\documentstyle[12pt]{article}
\newcommand{\Tr}[1]{\,{\rm Tr}\,#1\,}
\begin{document}

\begin{flushright}
{\small SMI-9-94 }
\end{flushright}

\vspace{2.5cm}
\begin{centering}

{\huge Generalized Pauli-Villars regularization for undoubled lattice
fermions}\\
\vspace{1cm}
{\large A.A.Slavnov\footnote{E-mail:$~~$ slavnov@qft.mian.su}\\
Steklov Mathematical Institute, Russian Academy of Sciences,\\
Vavilov st.42, GSP-1,117966, Moscow, Russia }\\
\end{centering}

\begin{abstract}

A manifestly gauge invariant formulation of chiral theories with fermions on
the lattice is developed. It combines SLAC lattice derivative \cite{DWY},
\cite{ACS}, \cite{S} and generalized Pauli-Villars regularization
\cite{FS}.  The theory is free of fermion doubling, requires only local
gauge invariant counterterms and produces correct results when applied to
exactly solvable two dimensional models.  \end{abstract}

\section {Introduction}

In this paper I discuss a possibility to describe chiral fermions on
the lattice
 in the framework of the SLAC formulation \cite{DWY}, \cite{ACS}, \cite{S}
supplemented by the additional Pauli-Villars (PV) type regularization
which allows to avoid the problems of the original SLAC model.

It is known that due to nonlocality of the SLAC action perturbation theory
suffers from the appearance of nonlocal counterterms and spurious infrared
divergencies \cite{KS}. The attempts to overcome this difficulties by making
a partial resummation \cite{R} when applied to solvable models lead to the
wrong physical spectrum \cite{BK}.

The origin of these difficulties is a singular behavior of the fermion
propagator
 and vertices near the edge of the Brillouin zone
$p_{ \mu} \sim \frac{\pi}{a} $. The contribution of this region produces
nonlocal counterterms and infrared divergencies. Essentially the same
problem is present in all other models proposed to eliminate fermion
doubling on the lattice \cite{W}, \cite{Sm}, \cite{Sw}. (For detailed
references see \cite{S1}, \cite{G}, \cite{P}). In particular in Wilson
formulation the contribution of the region near the edge of the Brillouin
zone produces gauge noninvariant counterterms. In other words a lattice
formulation is not a bona fide regularization for chiral fermions, and to
 get an invariant calculational scheme one needs to suppress a
contribution of momenta of the order of cut-off. In fact it is the same
problem which has to be solved if one wishes to construct a gauge
invariant regularization for the chiral fermions in the continuum theory.

Recently we proposed a manifestly invariant regularizaion for the
anomaly free
chiral models, in particular for the standard model \cite{FS} (see
also \cite{NN}).
Our procedure is a generalization of a Pauli-Villars regularization
which allows for an infinite number of auxiliary PV fields. It has been
shown that when applied to the chiral models on the lattice this procedure
leads to a gauge invariant continuum theory without fermion doubling both
in the case of Wilson fermions \cite{FS1} and in the Smit-Swift model
\cite{Sl}. However both these formulations have some drawbacks. It is a
lack of a gauge invariance for a finite lattice spacing in the first case,
and the necessity to introduce an additional Yukawa interaction in the
second case. It seems that the most appropriate way to implement this
regulaization in lattice models is to use the SLAC formulation.

We shall show that if one writes a lattice regularization for anomaly free
chiral gauge models following  \cite{DWY} \cite{ACS}, \cite{S} and
introduces simultaneously PV fields according to \cite{FS},the resulting
model is manifestly gauge invariant, has no fermion doubling, and does not
require nonlocal counterterms in perturbation theory. When applied to
solvable two dimensional models this procedure leads to a correct
continuum limit.

\section {Four dimensional $SO(10)$ model.}

In this section we consider the grand unified $SO(10)$ model. It is
worthwhile to emphasize that representations we consider are not real, and
the choice of $SO(10)$ is explained by the possibility to write all the
formulae in a compact form. We could consider equally well the standard
model or Weinberg-Salam model provided the lepton and quark
representations are choosen in such a way that anomalies are compensated.
The SLAC lattice action for the grand unified $SO(10)$ model looks as
follows:  \begin{equation} I = \sum_{k, \mu,x,y} \bar{ \psi}_+^k(x)
\gamma_{\mu}iD_{\mu}(x-y)P \exp \{i
 \sum_{z_{\mu}=x_{\mu}}^{y_{\mu}}gA_{\mu}(z) \} \psi_+^k(y) \label{1}
 \end{equation}
 Here $A_{\mu}=A_{\mu}^{ij} \sigma_{ij}$.  The matrices $ \sigma_{ij}$ are
the $SO(10)$ generators:$\sigma_{ij}=1/2[\Gamma_i,\Gamma_j]$, where
$\Gamma_i$ are Hermitian $32\times32$ matrices which satisfy the Clifford
algebra: $[\Gamma_i,\Gamma_j]=2\delta_{ij}$. The chiral $SO(10)$ spinors
$\psi_\pm=1/2(1\pm\Gamma_{11})\psi$, where $\Gamma_{11}=\Gamma_1\Gamma_2
\ldots\Gamma_{10}$, describe the 16-dimensional irreducible representation
of $SO(10)$ including quark and lepton fields. We assume also that the spinors
$\psi_\pm$ are left-handed $\psi_\pm=1/2(1+\gamma_5)\psi_\pm$. Index $k$
numerates different generations. $D_{\mu}(x)$ is the SLAC derivative
\begin{equation}
D_{\mu}(x) = \int_{\frac{- \pi}{ a}}^{\frac{\pi}{a}}
\frac{d^4k}{(2 \pi)^4}ik_{\mu} \exp \{ikx \}
\label{2}
\end{equation}
The action (\ref{1}) is manifestly invariant with respect to $SO(10)$ gauge
transformations and does not suffer from fermion doubling. However it is not
local which leads to difficulties we mentioned above.

To avoid these problems one modifies the action (\ref{1}) by adding the
analogous interaction of PV fields \cite{FS}, \cite{FS1}, \cite{Sl}
\begin{equation}
I \rightarrow I + I_{PV} \label{3}
\end{equation}
$$
I_{PV} = \sum_{r, \mu,x,y} \bar{ \psi}_r(x) \gamma_{\mu}iD_{\mu}(x-y)
P \exp \{i \sum_{z_{\mu}=x_{\mu}}^{y_{\mu}}gA_{\mu}(z) \} \psi_r(y) -
\frac{M_r}{2} \bar{\psi}_r(x)C_DC \Gamma_{11} \bar{\psi}_r^T +
$$
$$
+ \sum_{r, \mu,x,y} \bar{ \Phi}_r(x) \gamma_{\mu}i
\Gamma_{11}D_{\mu}(x-y) P \exp \{i
\sum_{z_{\mu}=x_{\mu}}^{y_{\mu}}gA_{\mu}(z) \} \Phi_r(y) - \frac{M_r}{2}
\bar{\Phi}_r(x)C_DC \bar{\Phi}_r^T + h.c.  $$ Here $ \psi_r$ are
anticommuting PV fields and  $\Phi_r$ are commuting PV fields, which
realize the reducible 32 dimensional representation of $SO(10)$, $C_D$ is
the usual charge conjugation matrix and $C$ is the $SO(10)$ conjugation
matrix $ \sigma_{ij}^TC=-C \sigma_{ij}$.

The action (\ref{3}) is manifestly invariant with respect to the $SO(10)$
gauge transformations. Now we shall show that it generates a perturbation
theory which in the limit $a \rightarrow 0$ results in a continuum
theory with
only local gauge invariant counterterms. Nonlocal counterterms do not appear.

The perturbative propagators look as follows
\begin{equation}
S_{\bar{\psi}^+ \psi^+} = \frac{\hat{P}}{P^2}  \label{4}
\end{equation}
\begin{equation}
S_{\bar{\psi}_r^+ \psi_r^+}= S_{\bar{\psi}_r^- \psi_r^-}=
S_{\bar{\Phi}_r^+ \Phi_r^+}=-S_{\bar{\Phi}_r^- \Phi_r^-}=
\frac {\hat{P}}{P^2 +M_r^2},
\label{5}
\end{equation}

\begin{equation}
S_{\bar{\psi}_r^- \bar{\psi}_r^+}= S_{\psi_r^+ \psi_r^-}=
S_{\bar{\Phi}_r^- \bar{\Phi}_r^+}=
S_{\Phi_r^+ \Phi_r^-}= \frac {M_r C_D C \Gamma_{11}}{P^2 + M_r^2 },
\label{6}
\end{equation}
where $P_{\mu}$ is the sawtooth function
\begin{equation}
P_{\mu}(p)=p_{\mu}-2m \frac{\pi}{a};\quad (2m-1)
\frac{\pi}{a}<p_{\mu}<(2m+1) \frac{\pi}{a}.  \label{7} \end{equation} The
expansion of the action (\ref{3}) in terms of $g$ generates the
interaction vertices with the increasing number of vector lines. (For
detailes see
\cite{KS}, \cite{R}). The three point vertex looks as follows
\begin{equation}
\Gamma^3=g \gamma_{\mu} \frac{P_{\mu}(p)-P_{\mu}(q)}{K_{\mu}(k)}
\sigma_{ij}, \label{8} \end{equation} with \begin{equation}
K_{\mu}(k)= \frac{ \exp[iak_{\mu}-1]}{ia}  \label{9}
\end{equation}
One sees that the interaction vertex (\ref{8}) is nonlocal, and it introduces
additional singularities into Feynman diagrams. Indeed, if $p< \frac{\pi}{a}$,
but $q> \frac{\pi}{a}$, the interaction vertex becomes $ \sim \frac{a^{-1}}
{K_{\mu}(k)}$, which results in the appearance of the nonlocal divergent
contributions \cite{KS}.

When the PV fields are introduced the contribution of the region
$p \sim a^{-1}$ is suppressed and nonlocal divergencies do not appear.

Let us consider for example the vacuum polarisation diagrams generated by the
vertex (\ref{8}). They look as follows:
\begin{equation}
\Pi_{\mu \nu}^{(\pm)r}= - \frac{g^2}{K_{\mu}K_{\nu}} \int_{\frac{- \pi}{a}}
^{\frac{\pi}{a}} \frac{d^4l}{(2 \pi)^4} \Tr [\sigma_{ij}(1 \pm
\Gamma_{11}) \sigma_{kl} ]  \times
\label{10}
\end{equation}
$$
\Tr [(1+ \gamma_5) \gamma_{\mu} (\hat{P}(l)+M_r)
\gamma_{\nu}(\hat{P}(l+p)+M_r) ]
\frac{[P_{\mu}(l)-P_{\mu}(l+p)][P_{\nu}(l+p)-
P_{\nu}(l)]}{[P^2(l)+M_r^2][P^2(l+p)+M_r^2]}.
$$
(We consider the contribution of the chirality preserving propagators
(\ref{4},
\ref{5})). The contribution of the original fermion is given by
$\Pi_{\mu \nu}^{+(0)}, (M_r=0)$.
Contribution of the bosonic fields differs by sign.

First of all we note that
\begin{equation}
\Tr [\sigma_{ij} \Gamma_{11} \sigma_{kl} ]=0.  \label{11}
\end{equation}
and therefore positive and negative "chirality" PV fields give the same
contribution to $\Pi_{\mu \nu}$. Secondly, the nonlocal contribution comes
from the region of integration where $\frac{\pi}{a}-|p_{\mu}| \leq
|l_{\mu}| \leq \frac{\pi}{a}$. If one chooses $M_r \ll a^{-1}$, one can
expand the integrand in this region in terms of $M_r$. Terms of zero order
produce the total contribution proportional to \begin{equation} k+2
\sum_r(-1)^rc_r  \label{12} \end{equation} Here $c_r$ is the number of
the PV fields with the mass $M_r$. The contributions of fermionic and
bosonic fields differ by sign. The factor $2$ arises because there are PV
fields of both chiralities, and the factor $k$ is due to the presence of
$k$ generations of original fermions. If the number of generations is even
one can always choose the coefficients $c_r$ in such a way that
\begin{equation}
k+2 \sum_r(-1)^rc_r=0  \label{13}
\end{equation}
Imposing the further PV conditions
\begin{equation}
\sum_rc_r(-1)^rM_r^2=0  \label{14}
\end{equation}
one can make the contribution of the domain $\frac{\pi}{a}-|p_{\mu}|
\leq |l_{\mu}| \leq \frac{\pi}{a}$  vanishing in the limit $a
\rightarrow0$.  In the remaining integral over $|l_{\mu}| \leq
\frac{\pi}{a}-|p_{\mu}|$, \begin{equation} P_{\mu}(l)-P_{\mu}(l+p) \sim
p_{\mu} \label{15} \end{equation} Therefore  the nonlocal factors
$K_{\mu}(p)$ are compensated and in the continuum limit one gets a
manifestly gauge invariant expression for the polarization operator.

The same arguments are obviously applicable to the diagrams including
chirality changing propagators (\ref{6}).

In the case of an odd number of generations the eq.(\ref{13}) cannot be
satisfied for any finite number of PV fields, as by construction the
coefficients
$c_r$ are integer. As was discussed in ref. \cite{FS}, \cite{FS1},
\cite{Sl} in this case the desired suppression of the region near the edge
of the Brillouin zone may be achieved if one introduces an infinite series
of PV fields with the masses $M_r=Mr$, $M(a) \rightarrow \infty$ when $a
\rightarrow 0$, $M \ll a^{-1}$. In this case after summation over $r$ one
gets the integrand \begin{equation} \sim \frac{\pi}{MR \sinh(\pi
RM^{-1})},\quad R^2=P_0^2(l)+P_1^2(l)+P_2^2(l)+ P_3^2(l)  \label{16}
\end{equation} which vanishes exponentially for $R \sim \frac{\pi}{a}$.
Therefore this region does not contribute and one can use the same
arguments as above to show that in the continuum limit one gets a
manifestly gauge invariant polarization operator and only usual local
gauge invariant counterterms are needed.

The proof for the diagrams with $3$ and $4$ external lines is
given in the same way.
The diagrams generated by the vertex function (\ref{8}) and the propagators
(\ref{4}),(\ref{5}) may be presented in the form
\begin{equation}
I_{\mu_1 \ldots \mu_n}= \frac{\tilde{I}_{\mu_1 \ldots \mu_n}}{K_{\mu_1}(p_1)
K_{\mu_2}(p_2) \ldots K_{\mu_n}(p_n)}   \label{21}
\end{equation}
where
\begin{equation}
\tilde{I}_{\mu_1 \ldots \mu_n}^{(\pm)r}=- g^n \int_{\frac{-
\pi}{a}}^{\frac{\pi}{a}}
 \frac{d^4l}{(2 \pi)^4}
\Tr [\sigma_{ij}(1 \pm \Gamma_{11}) \sigma_{kl} \ldots \sigma_{mn} ] \times
\label{17}
\end{equation}
$$
\Tr [(1+ \gamma_5) \gamma_{\mu_1} (\hat{P}(l)+M_r)\gamma_{\mu_2} \ldots
(\hat{P}(l-p_n)+M_r) ] \times
$$
$$
\frac{[P_{\mu_1}(l)-P_{\mu_1}(l+p_1)] \ldots
[P_{\mu_n}(l-p_n)-P_{\mu_n}(l)]}{[P^2(l)+M_r^2] \ldots [P^2(l-p_n)+M_r^2]}.
$$
Due to the properties of $\Gamma$ matrices $\Tr [\sigma_{i_1j_1} \Gamma_{11}
\ldots \sigma_{i_nj_n} ]=0$ for $n \leq 4$ and as before the contributions
of positive and negative chirality fermions are equal.
The integrals over $dl_i$ may be separated into two parts:
\begin{equation}
\int_{V_{in}}dl_i+\int_{V_{out}}dl_i, \quad
 V_{in}={|l_i|< \frac{\pi}{a}-|p_i|}, \quad V_{out}={|l_i|>
\frac{\pi}{a}-|p_i|} \label{18} \end{equation} Here $|p_i|= \sup |p_i^1+
\ldots p_i^l|, l=1, \ldots n-1$. If $l_{\mu}$ belongs to $V_{in}$ then the
factors $P_{\mu}(l+p^{1}+ \ldots p^l)-P_{\mu}(l+p^1+ \ldots p^{l-1})$ can
be replaced by $p_{\mu}^l$,and the integral over $V_{in}$ reduces in the
limit $a \rightarrow 0$ to the usual gauge invariant continuum expression.
Hence if we adjust the parameters of the PV fields in such a way that the
integral over $V_{out}$ vanishes in the continuum limit, our statement will
be proven.

In complete analogy with the analysis of the polarization operator in the
case of even number of generations of the original fields we may take a
finite number of PV fields with the parameters satisfying eqs. (\ref{13},
\ref{14}). Expanding the integrands in (\ref{17}) one sees that the leading
terms cancel and the remaining terms vanish in the continuum limit.

In the case of an odd number of generations one needs an infinite system of PV
fields. Taking the traces and separating tensorial structures in eq.
(\ref{17}) one can write for $ \tilde{I}_{\mu_1 \ldots \mu_n}$
\begin{equation}
\tilde{I}_{\mu_1 \ldots \mu_n} \sim \int_{\frac{- \pi}{a}}^{\frac{\pi}{a}}
\frac{d^4l}{(2 \pi)^4} \sum_{r=- \infty}^{+ \infty} \sum_{j=0}^{n-1}
\frac{A_j^{\mu_1 \ldots \mu_n}(l,Q,M_r)}{P^2(l+Q_j)+M^2r^2}, \quad Q_j=p_1+
\ldots + p_j.   \label{19} \end{equation} where $A_l^{\mu_1 \ldots \mu_n}$
is a polinomial in $M_r^2$. The summation over $r$ can be done explicitly
as in ref.\cite{FS}, \cite{FS1}. One gets \begin{equation}
I_{\mu_1 \ldots \mu_n} \sim \int_{\frac{- \pi}{a}}^{\frac{\pi}{a}}
\frac{d^4l}{(2 \pi)^4}\sum_{j=0}^{n-1} \frac{\tilde{A}_j^{\mu_1 \ldots
\mu_n}(l,Q,)}{MR \sinh (\pi RM^{-1})}   \label{20}
\end{equation}
In the region $V_{out}$ the integrand vanishes exponentially when $a
\rightarrow 0 $, and the integral over $V_{in}$ reduces to the gauge
invariant continuum expression.

To extend this proof to the higher order diagrams we firstly note that
contrary to
the statements which one can meet in the literature the model
(\ref{1}) generates a finite number of types of divergent diagrams. Indeed,
scaling the integration variables $l_{\mu} \rightarrow l_{\mu}a^{-1}$, in
eq.( \ref{17}) one would get for arbitrary diagram the highest degree of
divergency $ \sim a^{-4}$. However this estimation is too rude. Let us
show that
all the diagrams with $n \geq 5$ are convergent. To prove it we separate the
domain of integration over$dl_i$ as in eq.( \ref{18}). If $l_{\mu}$
belongs to
$V_{in}$ then the factors $P_{\mu}(l+p^{1}+ \ldots p^l)-P_{\mu}(l+p^1+
\ldots p^{l-1})$
can be replaced by $p_{\mu}^l$. Each such factor decreases the degree
of divergency
by one. The integral over $V_{out}$ is limited by the product of
independent external momenta $p_1p_2 \ldots p_{n-1}$. Indeed, the
integrand contains the factors
\begin{equation}
P_i(l+ap_1+ \ldots +ap_k)-P_i(l+ap_1+ \ldots +ap_{k+1}) \label{23}
\end{equation}
This factor can be different of $ap_{k+1}^i$ only on the interval of values
of $l^i$ which is equal to $ap_{k+1}^i$. Noting that the remaining terms
in the integrand are not singular at $p_{k+1}^i=0$, we conclude that
the integral
is limited by $F|ap_{k+1}^i|$ where $F$ is a polynomial over
$ap_{k+1}^i$. It follows that
the diagram with $n$ external lines contributes the factor
\begin{equation}
\tilde{I}_{\mu_1 \ldots \mu_n} \leq a^{-4}|ap_{\mu_1}| \ldots |ap_{\mu_n-1}|
\label{24}
\end{equation}
and therefore all the diagrams with $n \geq 5$ are convergent.

These reasonings are easily extended to the case when the diagram includes
the higher order vertices $ \Gamma_n(l,p_1 \ldots p_n)$. The vertices
$\Gamma_n$
are defined up to nonessential factors by the recurrent relations
\begin{equation}
\Gamma_{n+1}(l,p_1 \ldots p_{n+1} \sim -\Gamma_n(l,p_1 \ldots p_n)+
\Gamma_n(l+
p_{n+1},p_1 \ldots p_n) \label{24'}
\end{equation}
Let us consider some momentum $p_i^k$.It follows from eqs.( \ref{8},
\ref{24'})
that $\Gamma_n$ is the sum of $P_i$ depending on different arguments and all
the terms in $\Gamma_n$ may be separated into pairs whose arguments differ by
$p_i^k$. Therefore $\Gamma_n$ is proportional to $p_i^k$ everywhere
except for
the interval of the values of $l_i$ whose length is equal to $ap_i^k$.
In complete analogy with the case discussed above we conclude that the
eq.(\ref{24})
holds. Counting the number of independent momenta one sees that the
diagrams with
 $n>4$ are convergent. In the limit $a \rightarrow 0$ they coincide
with the gauge
invariant continuum expression. The corresponding diagrams with PV
fields vanish in
the continuum limit.

It is worthwhile to notice that these reasonings do not exclude
nonlocal singularities
for diagrams with $n<4$. For example in polarization operator there is
only one independent momentum and without PV regularization it
contains nonlocal
singularities.

This discussion shows also that spurious infrared divergencies which
could be caused by
the singularities of the vertex functions are in fact absent being
supressed according to
the eq.(\ref{24}).

The analysis of the diagrams with divergent subgraphs in particular
overlapping
divergencies requires more complicated machinery and has not yet been
done. We just
mention that this problem may be avoided by introducing an additional
higher derivative
regularization for Yang-Mills field, as it has been done in the refs.
\cite{FS1},
\cite{Sl}. With this regularization all the above results are
trivially extended to
arbitrary diagrams.

It completes the proof of the absence of nonlocal counterterms in our scheme.
Of course all the arguments given above used the weak coupling expansion and
it would be very important to verify them nonperturbatively. Having this in
mind we consider in the next section two dimensional chiral models.
\section {Two dimensional models}
In two dimensional models the procedure described above simplifies greatly
because the only divergent diagram is the second order polarization operator.
In the continuum Abelian model the polarization operator is the only nonzero
diagram and for that reason the anomaly free chiral models are
essentially
equivalent to the vectorial Schwinger model. It is not true in the nonabelian
case.

We start with the anomaly free chiral Schwinger model. The fermionic part of
the action is
\begin{equation}
I = \sum_{ \mu,x,y}[ \bar{ \psi}_+(x) \gamma_{\mu}iD_{\mu}(x-y)
 \exp \{iea \sum_{z_{\mu}=x_{\mu}}^{y_{\mu}}A_{\mu}(z) \} \psi_+(y) +
\label{25}
\end{equation}
$$
 \sum_{k=1,2} \bar{ \psi}_-^k(x) \gamma_{\mu}iD_{\mu}(x-y)
 \exp \{i \frac{e}{\sqrt{2}}a \sum_{z_{\mu}=x_{\mu}}^{y_{\mu}}A_{\mu}(z) \}
\psi_-^k(y)]
$$
where
\begin{equation}
D_{\mu}(x) = \int_{\frac{- \pi}{ a}}^{\frac{\pi}{a}}
\frac{d^2k}{(2 \pi)^2}ik_{\mu} \exp \{ikx \} \label{26}
\end{equation}
$ \psi_{\pm}= \frac{1}{2}(1 \pm \gamma_5) \psi$. The coupling constants of
the positive and negative chirality fermions are adjusted in such a way that
the $\gamma_{5}$ contributions cancel and the model is anomaly free.

To suppress the contribution of the region near the edges of the
Brillouin zone we
introduce the vectorial interaction of bosonic PV fields $\Phi$
\begin{equation}
I = \sum_{ \mu,x,y}[ \bar{ \Phi}(x) \gamma_{\mu}iD_{\mu}(x-y)
 \exp \{iea \sum_{z_{\mu}=x_{\mu}}^{y_{\mu}}A_{\mu}(z) \} \Phi(y) +
M \bar{ \Phi}(x) \Phi(x)]   \label{27}
\end{equation}
The regularized action $I_R=I+I_{PV}$ is invariant with respect to the gauge
transformations
\begin{equation}
\psi_+(x) \rightarrow \exp \{iea \xi (x) \} \psi_+(x); \quad \psi_-^k(x)
\rightarrow \exp \{i \frac{e}{\sqrt{2}}a \xi (x) \} \psi_-(x)  \label{28}
\end{equation}
$$
\Phi (x) \rightarrow \exp \{iea \xi (x) \} \Phi(x),
A_{\mu}(x) \rightarrow A_{\mu}(x)+a^{-1} [f(x_{\mu}+a)-f(x)].
$$

Consider firstly the vacuum polarization graph. The corresponding
amplitudes for each fermions are
$$
{}~
$$
\begin{equation}
\Pi_{\mu \nu}^{(\pm)a}= - \frac{{g_{\pm}}^2}{K_{\mu}K_{\nu}} \int_{\frac{-
\pi}{a}} ^{\frac{\pi}{a}} \frac{d^2l}{(2 \pi)^2} \Tr [ \frac{1}{2}(1 \pm
\gamma_5) \gamma_{\mu} \hat{P}(l) \gamma_{\nu}\hat{P}(l+p) ] \times
\label{29} \end{equation} $$ \frac{[P_{\mu}(l)-P_{\mu}
(l+p)][P_{\nu}(l+p)-P_{\nu}(l)]}{P^2(l)P^2(l+p)}.
$$
\begin{equation}
\Pi_{\mu \nu}^{(\pm)b}=- \frac{g_{\pm}^2 \delta_{\mu \nu}}{K_{\mu}K_{\nu}}
\int_{\frac{- \pi}{a}}^{\frac{\pi}{a}} \frac{d^2l}{(2 \pi)^2} \Tr [
\frac{1}{2}(1 \pm \gamma_5) \gamma_{\mu}(\hat{P}(l) ]
\frac{[-2P_{\mu}(l)+P_{\mu}(l+p)+ P_{\mu}(l-p)]}{P^2(l)}  \label{30}
\end{equation}
Here $\pm$ stands for the contributions of positive and negative chirality
fermions , and $g_+=e, g_-=\frac{e}{\sqrt{2}}$. The corresponding amplitudes
for the PV fields are
\begin{equation}
\Pi_{\mu \nu}^a= \frac{e^2}{K_{\mu}K_{\nu}} \int_{\frac{- \pi}{a}}
^{\frac{\pi}{a}} \frac{d^2l}{(2 \pi)^2} \Tr [ \gamma_{\mu}
(\hat{P}(l)+M_r) \gamma_{\nu}(\hat{P}(l+p)+M_r) ] \times
\label{31}
\end{equation}
$$
\frac{[P_{\mu}(l)-P_{\mu}
(l+p)][P_{\nu}(l+p)-P_{\nu}(l)]}{[P^2(l)+M_r^2][P^2(l+p)+M_r^2]}.
$$
\begin{equation}
\Pi_{\mu \nu}^b=2 \frac{e^2 \delta_{\mu \nu}}{K_{\mu}K_{\nu}} \int_
{\frac{- \pi}{a}}^{\frac{\pi}{a}} \frac{d^2l}{(2 \pi)^2} \Tr [
\gamma_{\mu}(\hat{P}(l)+M_r)]
\frac{[-2P_{\mu}(l)+P_{\mu}(l+p)+P_{\mu}(l-p)]} {P^2(l)+{M_r}^2}
\label{32} \end{equation} The complete contribution of the original
fermions is given by \begin{equation} \Pi_{\mu \nu}^{a+}+2\Pi_{\mu
\nu}^{a-}+\Pi_{\mu \nu}^{b+}+2\Pi_{\mu \nu}^{b-} \label{33}
\end{equation}
The terms proportional to $\gamma_5$ cancel and the sum (\ref{33}) coincides
up to the sign and mass terms with the PV amplitudes (\ref{31}, \ref{32}).
Following the procedure described in the preceeding section we can separate
the integration domain into $V_{in}$ and $V_{out}$. In the domain $V_{out}$,
$p \gg M$ and one can expand the integrands  in terms of $M$. The first terms
cancel and the next terms are majorated by $M^2a^2$. Therefore the integral
over $V_{out}$ vanishes in the continuum limit. The integral over $V_{in}$
in the limit $a \rightarrow 0$ coincides with the PV regularized continuum
theory. The same reasoning are applied when $l_0$ belongs to $V_{in}$ and
$l_1$ belongs to $V_{out}$ and visa versa.

The diagrams with more than two photon lines are analysed in the same way as
above. One can easily see that the infrared divergencies are absent. Indeed,
according to the estimate given above the integrals over fermion loops are
limited by the product of independent external momenta. Therefore the
infrared
singular factors are supressed. A straightforward power counting shows
that the
higher order sea-gull vertices give a contribution vanishing in the continuum
limit. For example the tadpole diagram produced by the second order vertex
contributes the factor
\begin{equation}
\Pi \sim \int_{- \frac{\pi}{a}}^{\frac{\pi}{a}}d^2k
\frac{P_{\mu}(p)-P_{\mu}(p+k)-
P_{\mu}(p-k)+P_{\mu}(p)}{K_{\mu}(k)K_{\mu}(k)}D_{\mu \mu}(k) \label{34}
\end{equation}
where $D_{\mu \mu}$ is the photon propagator. This expression is
infrared singular
for $p \sim \frac{\pi}{a}$. However it was shown in the preceeding section
that the integration over $p$ produces the factors $ \sim k_{\mu}^2$
compensating
this singularity. Therefore one can rescale the integration variables
$k_{\mu}
\rightarrow a^{-1}k_{\mu}$, which allows to estimate this integral when $a
\rightarrow 0$:
\begin{equation}
\Pi \sim a \int_{-\pi}^{\pi}d^2xF(x,ap) \label{35}
\end{equation}
where it is understood that $ \Pi$ is attached to some fermion loop with the
integration momentum $p$. According to the discussion above the integral is
convergent and therefore $ \Pi \rightarrow 0$ when $a \rightarrow 0$.
Analogous
reasonings show that all loops including internal photon lines vanish in the
continuum limit.  The only nonzero diagram is the photon
polarization operator which coincides with the continuum result:
\begin{equation}
\Pi_{\mu \nu}=- \frac{e^2}{\pi}(\delta_{\mu \nu}-
\frac{k_{\mu}k_{\nu}}{k^2}) \label{36} \end{equation}   Note that if
one would
apply to our model Rabin procedure without additional PV regularization
one would get for the mass gap the result which is the continuum one times
$ \sqrt{2}$ \cite{BK}. To avoid misanderstanding we emphasize that we did not
solve exactly the Schwinger model for a finite lattice spacing, but rather
demonstrated that the lattice approximation developed above generates in the
continuum limit the correct exact solution of the Schwinger model if the limit
is taken termwise in the weak coupling expansion.

\section {Discussion}

In this paper we showed that modification of the SLAC model by
introducing the
generalized PV regularization provides a manifestly gauge invariant lattice
description of anomaly free chiral models. Our model does not suffer from
fermion doubling and leads in the continuum limit to the usual gauge invariant
results with only local counterterms. When applied to the anomaly free chiral
Schwinger model it reproduces in the continuum limit the well known exact
solution.

{\bf Aknowlegements.} \\
I am grateful to J.Smit for helpful discussion.\\
This work was supported by International Science Foundation under grant
MNB000 and Russian Basic Research Fund under grant 94-01-00300a.  $$ ~ $$

\end{document}